**Strain engineering and one-dimensional organization of metal-insulator domains in single-crystal VO$_2$ beams**

J. Cao,[1,2] E. Ertekin,[3] V. Srinivasan,[3] W. Fan,[1,4] S. Huang,[1] H. Zheng,[2,5] J. W. L. Yim,[1,2] D. R. Khanal,[1,2] D. F. Ogletree,[2,6] J. C. Grossman,[3] and J. Wu[1,2,3] *

[1]Department of Materials Science and Engineering, University of California, Berkeley, Berkeley, CA 94720, USA

[2]Materials Sciences Division, Lawrence Berkeley National Laboratory, Berkeley, CA 94720, USA

[3]Berkeley Nanosciences and Nanoengineering Institute, University of California, Berkeley, Berkeley, CA 94720, USA

[4]Department of Thermal Science and Energy Engineering, University of Science and Technology of China, Hefei, China

[5]National Centre for Electron Microscopy, Lawrence Berkeley National Laboratory, Berkeley, CA 94720, USA

[6]Molecular Foundry, Lawrence Berkeley National Laboratory, Berkeley, CA 94720, USA

*To whom correspondence should be addressed. Email: wuj@berkeley.edu

**Spatial phase inhomogeneity at the nano- to microscale is widely observed in strongly-correlated electron materials. The underlying mechanism and possibility of artificially controlling the phase inhomogeneity are still open questions of critical importance for both the phase transition physics and device applications. Lattice strain has been shown to cause the coexistence of metallic and insulating phases in the Mott insulator $VO_2$. By continuously tuning strain over a wide range in single-crystal $VO_2$ micro- and nanobeams, here we demonstrate the nucleation and manipulation of one-dimensionally ordered metal-insulator domain arrays along the beams. Mott transition is achieved in these beams at room temperature by active control of strain. The ability to engineer phase inhomogeneity with strain lends insight into correlated electron materials in general, and opens opportunities for designing and controlling the phase inhomogeneity of correlated electron materials for micro- and nanoscale device applications.**

Correlated Electron Materials (CEMs) offer a wide spectrum of properties featuring various types of phase transitions, such as superconductivity, metal-insulator transition, and colossal magnetoresistance[1]. A spatial phase inhomogeneity or micro-domain structure is frequently observed in these materials[2], where multiple physical phases co-exist at the nano- to microscale at temperatures where a pure phase is expected. Despite decades of investigation, the question of whether the phase inhomogeneity is intrinsic or caused by external stimuli (extrinsic) still remains largely unanswered. This question not only plays a critical role in our understanding of the CEM physics, but also directly determines the spatial scale of CEM device applications.

Lattice strain, if tuned continuously, would be a sensitive means to shed light on the origin of the phase inhomogeneity. In contrast to conventional materials, where elastic deformation causes continuous, minor variations in material properties, lattice strain has profound influence on the electrical, optical, and magnetic properties of CEMs through coupling between the charge, spin, and orbital degrees of freedom of electrons[3]. If phase inhomogeneity is absent in strain-free, single-crystal specimens, but can be introduced and modulated by external strain, it would then be possible to eliminate or strain engineer the inhomogeneity and domains in CEMs for nanoscale device applications. Previous strain studies of CEMs have been limited to epitaxial thin films. Biaxial strain imposed from lattice mismatch with the substrate has been shown to remarkably enhance the order parameters in ferroelectric[4-6] and high-Tc superconducting epilayers[7]. In these films the lattice-mismatch strain distribution is complicated by misfit dislocations. In contrast, free-standing, single-crystal CEM nanostructures are dislocation-free, and can be subjected to coherent and continuously tunable external stress. CEM phase transitions and domain dynamics can then be explored through *in situ* microscopic experiments varying strain and temperature independently. Such an approach would enable, for the first time, probe of CEMs at the single domain level under continuous tuning of their lattice degree of freedom.

$VO_2$ is such a CEM that, in the strain-free state, undergoes a first-order metal-insulator phase transition (MIT) at $T_C^0 = 341$ K with a change in conductivity by several orders of magnitude. The MIT is accompanied by a structural phase transition from the high-temperature tetragonal phase (metallic, M) to the low-temperature monoclinic phase (insulating, I). Upon cooling through the MIT, the vanadium ions dimerize and these pairs tilt with respect to the tetragonal *c*-axis, causing the specimen to expand by $\varepsilon_0 \approx 1\%$ along the *c*-axis[8-10]. Along the tetragonal *a*- and *b*-axes, on the other hand, the lattice shrinks by 0.6 and 0.1%, respectively, causing a volume expansion of 0.3%[8-10]. The relationship between the MIT and the accompanied structural transition in $VO_2$ has been a topic of debate for decades[11-13]. As expected from the abrupt change in lattice constant upon the phase transition, a uniaxial compressive (tensile) stress along the tetragonal *c*-axis direction would drive the system toward the M (I) phase. In the stress-temperature phase diagram, the rate at which the transition temperature ($T_C$) is modified by the uniaxial stress ($\sigma$) is given by the Clausius-Clapeyron equation,

$$\frac{dT_C}{d\sigma} = \frac{\varepsilon_0 \cdot T_C^0}{\Delta H}, \qquad (1)$$

where $\Delta H$ is the latent heat of the transition. $dT_C/d\sigma$ was measured to be ~ 1.2 K/kbar for *c*-axis uniaxial stress[14]. As the volume change at the MIT is much weaker than the *c*-axis expansion, it is expected from the Clausius-Clapeyron equation that $T_C$ is much more sensitive to uniaxial stress than to hydrostatic pressure. This offers an efficient way to organize M-I domains in single-crystal $VO_2$ by imposing a uniaxial strain distribution. Such an effect has been shown in the pioneering work by Park's group, where single-crystal $VO_2$ nanobeams fully clamped on a $SiO_2$ surface spontaneously exhibit periodic M-I domains near $T_C^0$ owing to uniaxial strain imposed by elastic mismatch with the substrate[15,16]. The work was recently advanced by Cobden's group, where the fully clamped system was etched into an end-clamped configuration and interesting phase coexisting phenomena were observed[17]. In the present work, instead of passively observing the strain effect, we fully eliminate the substrate-imposed strain and then induce and continuously modulate the MIT by artificially stressing the $VO_2$. Single-crystal $VO_2$ micro and nanobeams were prepared with the length direction along the tetragonal *c*-axis (see Methods). We established coherent strain fields in the $VO_2$ beams by bending or applying uniaxial stress. The small width and the single-crystal nature of these beams allowed them to withstand an extraordinarily high uniaxial strain (> 2.5 %, as compared to <1% in bulk) without plastic deformation or fracture. The system responded to the strain field by self-organizing one-dimensionally micro- to nano-scale M-I domains along the beam length. Such an *active* and continuous control of the phase inhomogeneity opens possibilities for device applications of the MIT in $VO_2$. We generated a strain-temperature phase diagram for $VO_2$, and demonstrated the MIT in $VO_2$ at room temperature for the first time.

**Domain organization**

Figure 1**a-c** shows low- and high-resolution images of typical $VO_2$ beams. By adjusting the synthesis conditions, the beams can be grown with either a weakly coupled beam-substrate interface that slips to relieve stress, or a strongly coupled, clamped interface that pins the beam to the substrate. In the latter case thermal stress is imposed on the beams after cooling from the growth temperature to room temperature. Both types of beams were incorporated into four-probe devices using lithography. With increasing temperature, devices made from the un-stressed beams displayed a sharp drop of resistance at $T_C^0$ = 341 K (Fig. 1**d**, single-domain device). When heated through the transition temperature, the beam shows an abrupt change in brightness in white-light optical microscope images, from bright reflection in the low-temperature I phase to dark reflection in the high-temperature M phase. In contrast, the electrical resistance of the clamped beams decreases gradually across a wide phase-transition temperature range, showing an effective second-order phase transition. High-magnification optical imaging of the clamped beams revealed multiple M and I domains appearing during the transition, where dark domains nucleated in the bright phase and grew with increasing temperature, finally merging into a single M phase. A direct correlation between optical contrast and the electronic phases is therefore established, where bright and dark reflection indicates the I and M phase, respectively.

Clamped $VO_2$ beams with various widths (300 nm to 5 µm) displayed periodic M-I domains in high-resolution optical microscopy within the transition range (See Supplementary Information). Such a periodic domain pattern forms spontaneously as a result of the competition between strain energy in the elastically mismatched $VO_2$/substrate system and domain wall energy in the $VO_2$[16]. The period of the pattern is determined by the balance between the strain-energy minimization that favours small, alternating M-I domains and domain-wall energy minimization that opposes them[16,18]. The domain structure could not be resolved by optical microscopy on beams narrower than 300 nm, while beams wider than 5 µm showed irregular, two-dimensional domain texture possibly due to biaxial or non-uniform strain (as in the case of thin films). We focused our optical study on beams with widths between 1 and 2 µm, in which the micro-domains could be readily imaged and the local strain was easily controlled. In these beams a chain of M-I domains self-organize along the beam axis with a characteristic domain size comparable to the beam width.

To establish a wider range of coherent strain in the $VO_2$ beam, we bent non-clamped beams on the substrate by pushing part of the beam with a microprobe. A large compressive (tensile) strain results near the inner (outer) edge of the high-curvature regions of the bent beam. Figure 2**a** shows the development of an array of triangular domains along a bent $VO_2$ beam imaged at different temperatures. The bent beam was in I phase at room temperature. At elevated temperatures, sub-micron, periodic triangular M domains started to nucleate at the inner edge of the bent region where the strain was the most compressive. These domains continued to grow and expand with increasing temperature, while the triangular geometry and periodic arrangement were maintained. At $T \approx T_C^0 = 341$ K, the straight, strain-free part of the beam switched abruptly to the M phase as expected, while the bent part of the beam showed a nearly 50%-50% coexistence of M and I domains. These domains were highly periodic and triangular, with each triangles running through the entire width of the beam. As temperature was further increased, the M phase expanded toward the outer edge and finally completely eliminated the I phase at $T \approx 383$ K. Upon cooling (not shown), the domain evolution was reversed, with a ~ 10 K hysteresis from the heating process. Therefore, the domain organization at the microscale evolves from a periodic, nanoscale phase nucleation.

To quantitatively understand the strain stabilized M-I phase coexistence, we approximate the bent part of the beam with a constant curvature geometry where the strain varies linearly from compression to tension in the radial direction across a neutral plane. Assuming an equal Young's modulus for both the M and I phases ($Y$ = 140 GPa[19]), the stress is calculated from $\sigma = Y \cdot \varepsilon$ along the radial direction. The optically determined M-phase fraction $\eta$, at each radius, was measured from Fig.2**a** and plotted in $\sigma$ - $T_C$ phase space in Fig.2**b**. As expected, the system was in pure M phase ($\eta = 1$) at high temperatures and high compressive stresses, and in pure I phase ($\eta = 0$) at low temperatures and high tensile stresses. At intermediate temperatures and stresses, M and I phases coexisted with the spatial arrangement and relative fraction determined by energy minimization. According to Eq.(1), the boundary separating the M and I phases in the $\sigma$ - $T_C$ phase diagram in Fig.2**b** is directly related to the latent heat of the MIT. Fitting the experimental data using the upper and lower boundaries, we obtain a latent heat of $\Delta H$ between 1200 and 950 Cal/mol. This value is consistent with $\Delta H$ of 1025 Cal/mol reported for bulk $VO_2$[20,21]. We note that these values are nearly four times higher than the latent heat calculated from the $\sigma$ - $T_C$ data reported in the Cobden work[17].

Assuming a linear M-I phase boundary, the $\sigma$ - $T_C$ phase diagram predicts that a compressive strain of ~ 2.2% would be sufficient to drive $VO_2$ from I to M phase at room temperature. To test this prediction, we applied compressive stress directly along the length of a $VO_2$ beam clamped onto a soft substrate[22] and investigated how $\eta$ changed with external stress (see Methods). With increasing compressive stress along the beam axis, periodic M phase (dark domains) emerged out of the I phase (bright part) and gradually expanded, eventually merging to form a pure M phase beam. The $VO_2$ beam was monitored optically and $\eta$ was measured as a function of the total strain. As shown in Fig.2**c**, the beam remained entirely insulating ($\eta$=0) until stressed to a total strain of approximately $\varepsilon \approx -1.9\%$, then entered a strain regime where periodic M and I domains coexisted, ultimately reaching a full M state ($\eta$=1) at $\varepsilon \approx -2.1\%$. Upon releasing the strain from the M state, the $\eta$ - $\varepsilon$ curve showed a hysteretic behaviour and the beam returned to a full I state at $\varepsilon \approx -1.8\%$. As the resistance of the $VO_2$ beams changes by several orders of magnitude across the MIT, this actively controlled, room-temperature phase transition can be used as a "strain-Mott" transistor. Very recently, Joule heating-induced MIT in single $VO_2$ nanowires was used to achieve a novel type of gas sensors[23]. The nanowire was self-heated into the M phase when the bias voltage exceeded a threshold ($V_{th}$). The sensitivity and selectivity to different pressures and species of gas molecules came from their different thermal conductivity to dissipate the Joule heat away from the nanowire. However, a large $V_{th}$, thus a high operation power, was typically needed to achieve such device function[23]. The strain sensitivity of the MIT provides a strategy to drastically reduce the operation power and hence potentially increase the sensitivity and life time of these devices. In Fig.2**d** we show such effect observed from a self-heated VO$_2$ device under uniaxial compression at room temperature. Over the range of compression, $V_{th}$ and threshold current were reduced by a factor of 6 and 5, respectively; consequently, the operation power was reduced by a factor of 30. Further compression directly drove the device to M state without need for Joule heating, which can be seen from the trend of decreasing resistance of the I-phase at $V < V_{th}$. Active and extensive control of strain in these nano/microbeams thus offers a new "knob" to tune the MIT for novel or improved device applications.

**Phase field modelling**

To understand if the domain patterns self-organized along the beams are to be expected given the material parameters of VO$_2$ and the geometric dimensions of the beams, we implemented a two-dimensional (2D) phase field model, where the total energy $F(\phi)$ is equal to the sum of bulk thermodynamic energy, interfacial (domain wall) energy, and strain energy,

$$F(\phi) = \int \left[ f(\phi) + \frac{\beta^2}{2} |\nabla \phi|^2 + \frac{1}{2} C_{ijkl} \left( \varepsilon_{ij} - \varepsilon_{ij}^T \right) \left( \varepsilon_{kl} - \varepsilon_{kl}^T \right) \right] dA \cdot \qquad (2)$$

The parameter $\phi$ denotes the phase, in which $\phi = 0$ corresponds to M and $\phi = 1$ corresponds to I. A double well potential, $f(\phi)$, describes the relative thermodynamic energy of the M and I phases and is temperature dependent. The second term reflects the interfacial energy. The last term is the elastic energy where $C$ is the elastic modulus tensor, $\varepsilon$ is the strain, and $\varepsilon^T$ is the lattice mismatch between the two phases: $\varepsilon_{xx}^T$ varies smoothly from $\varepsilon_0 = 0$ to 1% as the phase varies from $\phi = 0$ to 1, and $\varepsilon_{yy}^T$ and $\varepsilon_{xy}^T$ are 0. The parameters used in the modelling are typical for bulk VO$_2$[16,19]. In the simulations, the phase distribution evolves from an initial random phase distribution (Fig. 3**a**) according to Cahn-Allen dynamics under boundary conditions corresponding to a uniform beam bending. Evolution of the phases and accompanied strain field relaxation were assembled into videos and shown in the Supplementary Information.

At $T = T_C^0$, the equilibrium phase distribution exhibits a periodic, triangular domain pattern as shown in Fig. 3**b**, demonstrating quantitatively that realistic geometric, elastic, and interfacial energy parameters give rise to the patterns observed in the experiment. This pattern nearly completely relieves the strain energy in the bent beam, with some remnant strain at the triangular tips (Fig.3**c**). In between the upper and lower edges, the ratio of metal/insulator phase varies linearly for optimal strain energy relief, corresponding exactly to the initial linear variation of strain. The period varies for different interfacial energy density and elastic constants. Specifically, the period is determined by competing effects of strain energy relaxation and interfacial energy minimization: smaller period results in more effective strain energy relief, but at the cost of introducing more interfacial area. This effect is commonly observed in, for instance, spinodally decomposing systems in which strain stabilizes periodic microstructures[24]. Lastly, we illustrate in Fig.3 the equilibrium phase distributions for systems that are slightly below or above $T_C^0$, which also agree well with the experimental observation in Fig.2**a**.

It is intriguing to compare the coexistence of M and I phases in the VO$_2$ beams to the phase inhomogeneity observed in CEM thin films, such as Mott insulators[25], colossal magnetoresistive manganites[2,26-32] and high-$T_C$ superconductors[33]. Specifically in VO$_2$, using scanning near-field infrared microscopy, Qazilbash *et al* recently observed an M-I phase coexistence within a range near $T_C^0$ in polycrystalline VO$_2$ films grown on Al$_2$O$_3$ substrates[25]. Upon heating across $T_C^0$ the M phase grows out of the I phase in a random, percolative manner with the size of M and I domain ranging from the nano to micro-scale. In contrast, in our single-crystal VO$_2$ beams, not surprisingly, the M and I phases coexist and evolve from the nano to micro-scale in an orderly fashion in response to the continuous tuning of coherent strain.

**Domain manipulation**

The sensitivity of the electronic phases to local strain allows one to manipulate and engineer the functional domains through external stress. Figure 4 shows various domain patterns stabilized in one VO$_2$ beam by bending different parts of the beam. Initially the beam was strain-free, and therefore in pure I phase at room temperature and pure M phase at 343 K. When the beam was locally bent at 343 K, an array of triangular I domains were created in the high curvature region in response to the local tensile strain. These domains were highly mobile and could be driven to different location along the beam by slight modifications of the bending geometry. Strain manipulation of M-I domains along a VO$_2$ beam was recorded as a video and is included in the Supplementary Information. In all cases, the spatial inhomogeneity and arrangement of the M and I phases could be understood from the total energy minimization of the system. The active control of the phase transition by strain offers a new approach to directly probe sub-domain behaviour of VO$_2$ in remote phase space, such as the M$_2$ phase[8] and critical point of the MIT[34].

**Conclusions**

In summary, lattice strain was used to actively control the phase inhomogeneity in the Mott insulator VO$_2$. We engineered metal-insulator domains along single-crystal VO$_2$ nano- and microbeams by uniaxial external stress. The domain structure was determined by competition between elastic deformation, thermodynamic, and domain wall energies in this coherently strained system. A uniaxial compressive strain of ~ 1.9% was able to stabilize coexisting metal and insulator domains at room temperature. The evolution of the electronic phases in VO$_2$ at the nano to micro-scale sheds light on the origin of the widely observed phase inhomogeneity in strongly correlated electron materials in general. The ability to engineer phase inhomogeneity and phase transition with strain demonstrated in this work opens opportunities for designing and controlling functional domains of VO$_2$ for micro- and nanoscale device or sensor applications. As distinctly different physical and chemical properties are associated with these phases, interfacing strain-engineered VO$_2$ with other molecular, nano- or polymeric materials may provide new assembly strategies to achieve collective and externally tunable functionalities.

**Acknowledgements**

This work was supported in part by National Science Foundation under Grant No. EEC-0425914, and in part by the Laboratory Directed Research and Development Program of Lawrence Berkeley National Laboratory (LBNL) under the Department of Energy Contract No. DE-AC02-05CH11231. Portions of this work were performed at the Molecular Foundry and at the National Centre for Electron Microscopy, LBNL.

**Methods**

Single-crystalline VO$_2$ beams were grown using a vapour transport method reported previously (See Supplementary Information) and characterized using scanning electron microscopy (SEM), transmission electron microscopy (TEM) and selected area electron diffraction[16,35]. These VO$_2$ beams grow along the tetragonal *c*-axis with {110} planes as the bounding side faces[35]. The width of these beams varies from 50 nm to a few microns and their length ranges from tens to a few hundred microns. In the experiment of applying compressive stress to single VO$_2$ beams, single-crystal VO$_2$ beams (typical length ~ 100 μm, width 0.5 ~2 μm) were transferred to a polycarbonate or Kapton substrate (thickness ~ 1 mm). A set of metal contacts were patterned using lithography and deposited using sputtering (15nm Cr and 400nm Au) for electromechanical measurements. Epoxy was used to bury and bond the $VO_2$ beam onto the substrate, and cured immediately at 390 K for 30 minutes. This step assured that the $VO_2$ beam was clamped onto the substrate when it was in M phase, and therefore when the $VO_2$ was cooled down to I phase at room temperature, a spontaneous compressive strain of $\varepsilon_0 \approx 1\%$ (due to the MIT) and another compressive strain of ~ 0.4% (due to thermal expansion mismatch with the polycarbonate substrate) were frozen in the system. By three-point concave bending the polycarbonate or Kapton substrate along the length direction of the $VO_2$ beam, further uniaxial compressive stress was added to the $VO_2$ beam. A strain gauge was glued in the centre of the substrate and used to monitor the additional strain imposed by the three-point bending.

In the phase field simulations, the total energy arises from thermodynamic bulk energy, interfacial energy, and strain energy. The thermodynamic energy is represented by a double-well potential; the relative depth of the minima is shifted to represent effects of temperature. The interfacial energy, isotropic here, is obtained by gradient terms which are non-zero in the transition regions. The phase field evolves according to Cahn-Allen dynamics for non-conserved order parameters, and is solved on a two-dimensional grid using finite differences that are second order accurate in space and first order accurate in time. A simple relaxation (Gauss-Seidel) method is implemented to find the displacement fields, strain, and stress by solving the equations of mechanical equilibrium with fixed displacement boundary conditions. The system considered corresponds to a beam of length 5 μm and width 1 μm, which is bent so that the strain on the upper and lower edge is 0.5% and -0.5%, respectively.

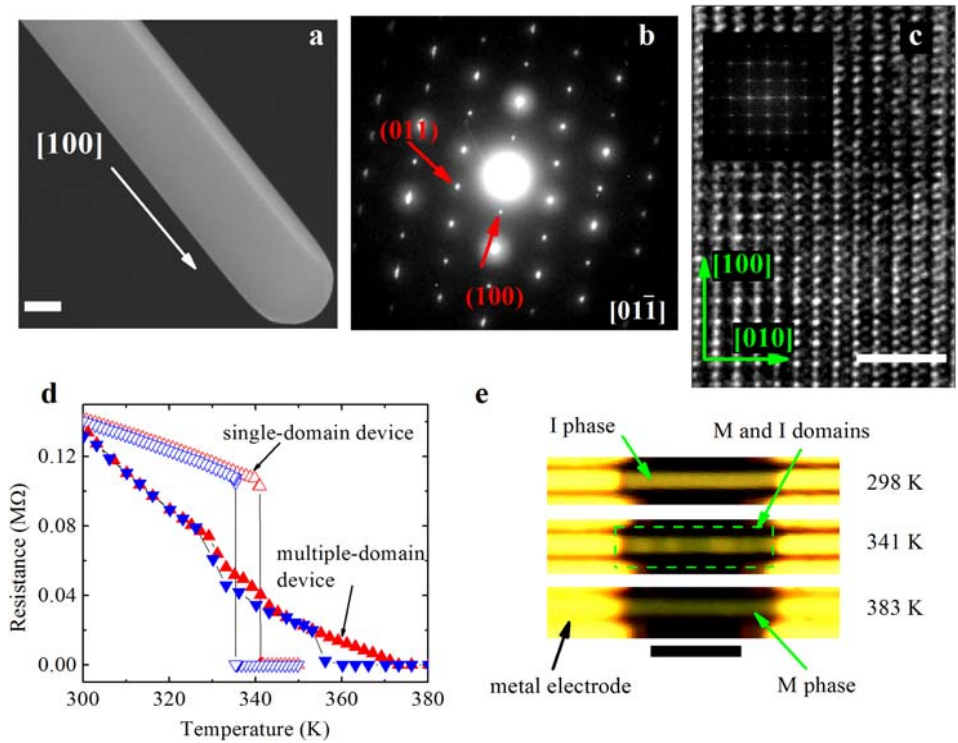

**Figure 1 Structural and electrical characterization of VO$_2$ beams. a,** SEM of a VO$_2$ beam grown along the monoclinic [100] direction (tetragonal *c*-axis) showing the facet surface. **b,** Selected area electron diffraction pattern for a 100 nm-wide nanobeam indexed using a [01$\bar{1}$] zone axis. **c,** High-resolution TEM image of a VO$_2$ nanobeam. The inset shows the corresponding FFT pattern indexed to monoclinic VO$_2$ with a [001] zone axis. **d,** Resistance of VO$_2$ beams measured in four-probe geometry as a function of temperature. The free-standing VO$_2$ beam shows single domain behaviour, whereas the clamped VO$_2$ beam shows multiple domains during the transition. (red is heating, blue cooling) **e,** Optical images of the multiple-domain device taken at 298, 341, and 383 K. The top and bottom images show the pure insulating and metallic phases, respectively. The middle image shows the coexistence metallic and insulating domains in the device. The scale bars in **a**, **c**, **e** are 100 nm, 2 nm, and 5 μm, respectively.

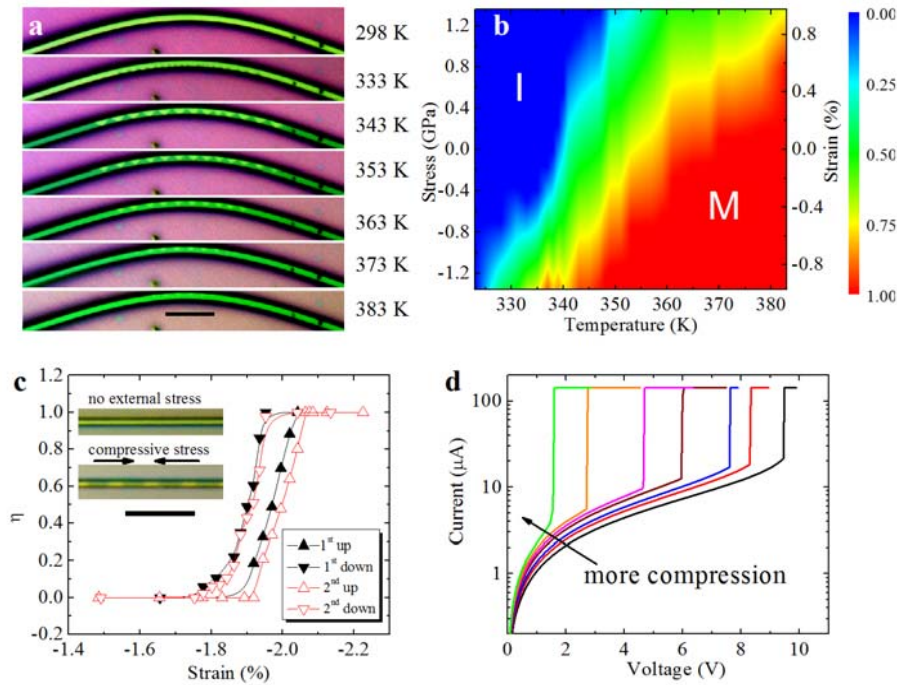

**Figure 2 Strain-induced MIT along VO$_2$ beams. a**, Optical images of an array of triangle M-I domain nucleated and co-stabilized by tensile and compressive strain during heating. Bright is I phase and dark is M phase. A colour-inverted image is shown in the Supplementary Information for better contrast between these two phases. **b**, Uniaxial stress vs. temperature phase diagram of extracted from the bent beam. The colour represents the optically determined fraction of M phase (η). **c**, Uniaxial compression reversibly induces MIT at room temperature in clamped VO$_2$ beams. The inset shows representative optical images of M-I domains along a stressed beam. **d**, Room-temperature I-V characteristic of a VO$_2$ beam under axial compression, showing Joule-heating induced MIT with threshold voltage and current drastically reduced by the external compression. The experiment was carried out in ambient with a current compliance of 150μA. The scale bars in **a** and **c** are both 10 μm.

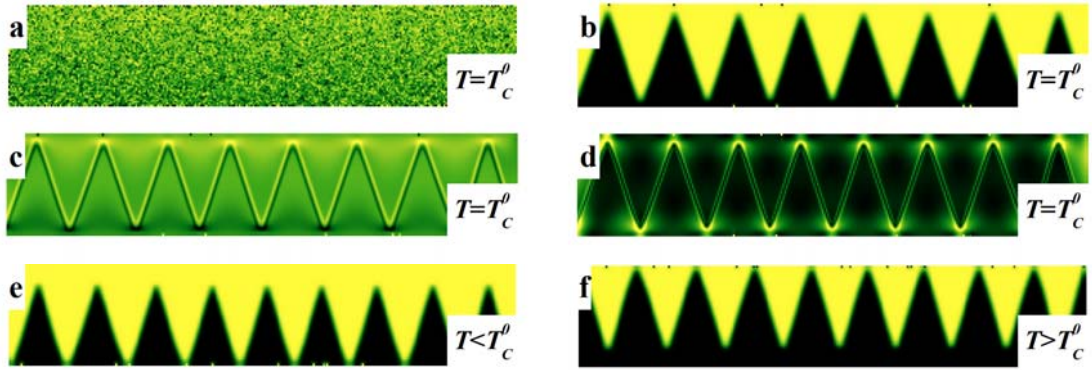

**Figure 3. Phase field modelling of domain formation in a bent VO$_2$ beam. a**, Initial M and I phase randomly distributed. Temperature is $T_C^0$, the natural MIT temperature. **b,** Equilibrium phase distribution showing M and I domain array self-organized at $T_C^0$. **c**, Equilibrium strain ($\varepsilon_{xx}$) distribution at $T_C^0$. **d**, Equilibrium strain energy density distribution. **e&f,** Equilibrium phase distribution showing more (less) I phase for Temperature lower (higher) than $T_C^0$. In **a, b, e, &f**, the yellow and dark green colours represent I and M phase, respectively. In **c**, the yellow and dark green colours denote the maximum tensile and maximum compressive strain, respectively. In **d,** the yellow and dark green colours denote the highest and lowest strain energy density, respectively.

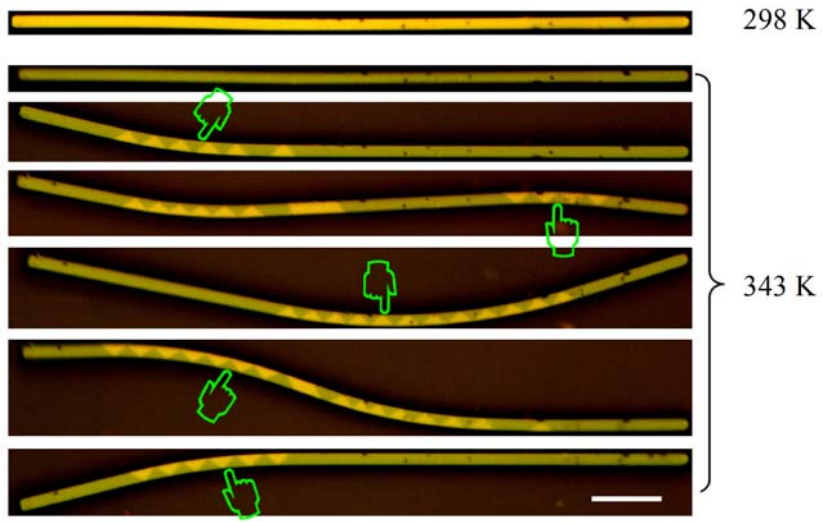

**Figure 4 Strain engineering domains in a VO$_2$ beam.** The un-bent beam was purely insulating (bright, top image) at 298 K and purely metallic (dark, second image) at 343 K. A tungsten needle (denoted by the hand symbol) was used to push-bend the beam, which created I domain arrays in the strained regions. The scale bar is 10 μm.